\documentstyle[12pt]{article}

\def\thefootnote{\fnsymbol{footnote}}
\def\ref#1{$^{#1)}$}

\def\journal{\topmargin .3in	\oddsidemargin .5in
	\headheight 0pt	\headsep 0pt
	\textwidth 5.625in 
	\textheight 8.25in 
	\marginparwidth 1.5in
	\parindent 2em
	\parskip .5ex plus .1ex		\jot = 1.5ex}
\journal

\catcode`\@=11
\def\marginnote#1{}
\newcount\hour
\newcount\minute
\newtoks\amorpm
\hour=\time\divide\hour by60
\minute=\time{\multiply\hour by60 \global\advance\minute by-\hour}
\edef\standardtime{{\ifnum\hour<12 \global\amorpm={am}%
	\else\global\amorpm={pm}\advance\hour by-12 \fi
	\ifnum\hour=0 \hour=12 \fi
	\number\hour:\ifnum\minute<10 0\fi\number\minute\the\amorpm}}
\edef\militarytime{\number\hour:\ifnum\minute<10 0\fi\number\minute}
\def\draftlabel#1{{\@bsphack\if@filesw {\let\thepage\relax
   \xdef\@gtempa{\write\@auxout{\string
      \newlabel{#1}{{\@currentlabel}{\thepage}}}}}\@gtempa
   \if@nobreak \ifvmode\nobreak\fi\fi\fi\@esphack}
	\gdef\@eqnlabel{#1}}
\def\@eqnlabel{}
\def\@vacuum{}
\def\draftmarginnote#1{\marginpar{\raggedright\scriptsize\tt#1}}
\def\draft{\oddsidemargin -.5truein
	\def\@oddfoot{\sl preliminary draft \hfil
	\rm\thepage\hfil\sl\today\quad\militarytime}
	\let\@evenfoot\@oddfoot	\overfullrule 3pt
	\let\label=\draftlabel
	\let\marginnote=\draftmarginnote
   \def\@eqnnum{(\theequation)\rlap{\kern\marginparsep\tt\@eqnlabel}%
\global\let\@eqnlabel\@vacuum}  }
\def\preprint{\twocolumn\sloppy\flushbottom\parindent 2em
	\leftmargini 2em\leftmarginv .5em\leftmarginvi .5em
	\oddsidemargin -.5in	\evensidemargin -.5in
	\columnsep .4in	\footheight 0pt
	\textwidth 10in	\topmargin  -.4in
	\headheight 12pt \topskip .4in
	\textheight 7.1in \footskip 0pt
	\def\@oddhead{\thepage\hfil\addtocounter{page}{1}\thepage}
	\let\@evenhead\@oddhead	\def\@oddfoot{}	\def\@evenfoot{} }
\def\numberbysection{\@addtoreset{equation}{section}
	\def\theequation{\thesection.\arabic{equation}}}
\def\underline#1{\relax\ifmmode\@@underline#1\else
	$\@@underline{\hbox{#1}}$\relax\fi}
\def\titlepage{\@restonecolfalse\if@twocolumn\@restonecoltrue\onecolumn
     \else \newpage \fi \thispagestyle{empty}\c@page\z@
	\def\thefootnote{\fnsymbol{footnote}} }
\def\endtitlepage{\if@restonecol\twocolumn \else \newpage \fi
	\def\thefootnote{\arabic{footnote}}
	\setcounter{footnote}{0}}  
\catcode`@=12
\relax
\def\figcap{\section*{Figure Captions\markboth
	{FIGURECAPTIONS}{FIGURECAPTIONS}}\list
	{Figure \arabic{enumi}:\hfill}{\settowidth\labelwidth{Figure 999:}
	\leftmargin\labelwidth
	\advance\leftmargin\labelsep\usecounter{enumi}}}
 \relax
\def\tablecap{\section*{Table Captions\markboth
	{TABLECAPTIONS}{TABLECAPTIONS}}\list
	{Table \arabic{enumi}:\hfill}{\settowidth\labelwidth{Table 999:}
	\leftmargin\labelwidth
	\advance\leftmargin\labelsep\usecounter{enumi}}}
 \relax
\def\reflist{\section*{References\markboth
	{REFLIST}{REFLIST}}\list
	{[\arabic{enumi}]\hfill}{\settowidth\labelwidth{[999]}
	\leftmargin\labelwidth
	\advance\leftmargin\labelsep\usecounter{enumi}}}
 \relax
\makeatletter
\newcounter{pubctr}
\def\publist{\@ifnextchar[{\@publist}{\@@publist}}
\def\@publist[#1]{\list
	{[\arabic{pubctr}]\hfill}{\settowidth\labelwidth{[999]}
	\leftmargin\labelwidth
	\advance\leftmargin\labelsep
	\@nmbrlisttrue\def\@listctr{pubctr}
	\setcounter{pubctr}{#1}\addtocounter{pubctr}{-1}}}
\def\@@publist{\list
	{[\arabic{pubctr}]\hfill}{\settowidth\labelwidth{[999]}
	\leftmargin\labelwidth
	\advance\leftmargin\labelsep
	\@nmbrlisttrue\def\@listctr{pubctr}}}
 \relax
\makeatother
\catcode`\@=11
\def\section{\@startsection {section}{1}{0pt}{-3.5ex plus -1ex minus
 -.2ex}{2.3ex plus .2ex}{\raggedright\large\bf}}
\catcode`\@=12
\newskip\humongous \humongous=0pt plus 1000pt minus 1000pt

\newif\ifdtup

\def\oldreffmt#1{\rlap{[#1]} \hbox to 2\parindent{}}

\def\figfmt#1{\rlap{Figure {#1}} \hbox to 1in{}}

\def\ie{\hbox{\it i.e.}}

\def\VEV#1{\left\langle #1\right\rangle}
\let\vev\VEV

\def\abs#1{\left| #1\right|}

\def\half{{1\over 2}}
\def\beq{\begin{equation}}
\def\eeq{\end{equation}}

\def\bea{\begin{eqnarray}}

\def\eea{\end{eqnarray}}
\def\ap#1#2#3{           {\it Ann. Phys. (NY) }{\bf #1}, #2 (19#3)}

\def\np#1#2#3{           {\it Nucl. Phys. }{\bf #1}, #2 (19#3)}
\def\pl#1#2#3{           {\it Phys. Lett. }{\bf #1}, #2 (19#3)}
\def\pr#1#2#3{           {\it Phys. Rev. }{\bf #1}, #2 (19#3)}

\def\prl#1#2#3{          {\it Phys. Rev. Lett. }{\bf #1}, #2 (19#3)}

\catcode`\@=11
\def\eqnarray{\stepcounter{equation}\let\@currentlabel=\theequation
\global\@eqnswtrue
\global\@eqcnt\z@\tabskip\@centering\let\\=\@eqncr
\gdef\@@fix{}\def\eqno##1{\gdef\@@fix{##1}}%
$$\halign to \displaywidth\bgroup\@eqnsel\hskip\@centering
  $\displaystyle\tabskip\z@{##}$&\global\@eqcnt\@ne
  \hskip 2\arraycolsep \hfil${##}$\hfil
  &\global\@eqcnt\tw@ \hskip 2\arraycolsep $\displaystyle\tabskip\z@{##}$\hfil
   \tabskip\@centering&\llap{##}\tabskip\z@\cr}

\def\@@eqncr{\let\@tempa\relax
    \ifcase\@eqcnt \def\@tempa{& & &}\or \def\@tempa{& &}
      \else \def\@tempa{&}\fi
     \@tempa \if@eqnsw\@eqnnum\stepcounter{equation}\else\@@fix\gdef\@@fix{}\fi
     \global\@eqnswtrue\global\@eqcnt\z@\cr}

\catcode`\@=12
\font\tenbifull=cmmib10 
\font\tenbimed=cmmib10 scaled 800
\font\tenbismall=cmmib10 scaled 666
\def\boldlambda{\fam=9{\mathchar"7115 } }

\def\boldnu{\fam=9{\mathchar"7117 } }

\def\boldLambda{\fam=6{\mathchar"7003 } }

\relax

\def\roughly#1{\mathrel{\raise.3ex\hbox{$#1$\kern-.75em%
    \lower1ex\hbox{$\sim$}}}}

\def\gsim{\roughly>}

\newcommand{\kapo}{\kappa_1}
\newcommand{\kapt}{\kappa_2}

\newcommand{\boldl}{\boldlambda}
\newcommand{\boldL}{\boldLambda}

\newcommand{\ffxv}{\vev{45_X}  }

\newcommand{\bfu}{ {\bf U} }
\newcommand{\bfd}{ {\bf D} }
\newcommand{\bfe}{ {\bf E} }

\newcommand{\otsbar}{\overline{126} }
\newcommand{\phione}{\phi_1}
\newcommand{\phitwo}{\phi_2}

\newcommand{\sixttw}{16_2}
\newcommand{\sixtth}{16_3}

\newcommand{\etab}{\eta_b}

\newcommand{\alphas}{\alpha_S}

\newcommand{\gu}{G_U}
\newcommand{\gd}{G_D}
\newcommand{\gee}{G_E}
\newcommand{\gon}{g_1}
\newcommand{\gtw}{g_2}
\newcommand{\gth}{g_3}

\newcommand{\be}{\begin{equation}}
\newcommand{\ee}{\end{equation}}

\newcommand{\au}{A_U}
\newcommand{\ad}{A_D}

\newcommand{\ba}{\begin{eqnarray*}}
\newcommand{\ea}{\end{eqnarray*}}

\newcommand{\ban}{\begin{eqnarray}}
\newcommand{\ean}{\end{eqnarray}}

\newcommand{\vcb}{V_{cb}}
\newcommand{\vcbo}{V_{cb}^0}
\newcommand{\thetut}{\theta_{\mu\tau}}
\newcommand{\thetuto}{\theta_{\mu\tau }^0}

\newcommand{\mg}{M_{GUT}}
\newcommand{\mz}{m_Z}
\newcommand{\mpp}{M_P}

\newcommand{\mt}{m_t}
\newcommand{\mb}{m_b}
\newcommand{\mtau}{m_\tau}

\newcommand{\tb}{\tan \beta}

\newcommand{\cb}{\cos \beta}

\newcommand{\htee}{h_t}
\newcommand{\hb}{h_b}
\newcommand{\htau}{h_\tau}

\newcommand{\ftp}{f_t}
\newcommand{\fbp}{f_b}
\newcommand{\ftaup}{f_\tau}

\newcommand{\zf}{16\pi^2}

\newcommand{\fbar}{\overline{5}}

\newcommand{\numt}{\nu_\mu-\nu_\tau}

\begin{document}
\begin{titlepage}
\begin{center}
\today     \hfill    LBL-33936 \\
          \hfill     hep-ph/9307275\\

\vskip .5in

{\large \bf $SO(10)$ Operator Analysis For $\nu_\mu \nu_\tau$ Oscillations.}
\footnote{This work was supported in part by the Director, Office of
Energy Research, Office of High Energy and Nuclear Physics, Division of
High Energy Physics of the U.S. Department of Energy under Contract
DE-AC03-76SF00098 and in part by the National Science Foundation under
grant PHY90-21139.}

\vskip .5in
H.-C. Cheng\\
M.S. Gill\\
and\\
L.J. Hall\\[.5in]

{\em  Department of Physics\\
      University of California\\
      and\\
      Theoretical Physics Group\\
      Physics Division\\
      Lawrence Berkeley Laboratory\\
      1 Cyclotron Road\\
      Berkeley, California 94720}
\end{center}

\vskip .5in

\newpage
\begin{abstract}
In grand unified theories the flavor mixing angles of leptons are related to
those of quarks. In this paper we study SO(10) theories with a precise group
theoretic relation at the GUT scale for mixing between the heaviest two
generations: $\thetuto = \kappa \abs{\vcbo}$. A comprehensive operator search
yields all possible cases where $\kappa$ is a group theory Clebsch. The
resulting predictions for $\nu_\mu \nu_\tau$ oscillations are scaled from the
grand to the weak scales.   We find that all but one of the models which have
such a relationship between $\thetuto$ and $\vcbo$,  and are not already
excluded, will be probed by the CHORUS and NOMAD experiments. A more precise
mixing measurement, for example by the proposed P803 experiment, could
distinguish between the models.
\end{abstract}
\end{titlepage}
\renewcommand{\thepage}{\roman{page}}
\setcounter{page}{2}
\mbox{ }

\vskip 1in

\begin{center}
{\bf Disclaimer}
\end{center}

\vskip .2in

\begin{scriptsize}
\begin{quotation}
This document was prepared as an account of work sponsored by the United
States Government.  Neither the United States Government nor any agency
thereof, nor The Regents of the University of California, nor any of their
employees, makes any warranty, express or implied, or assumes any legal
liability or responsibility for the accuracy, completeness, or usefulness
of any information, apparatus, product, or process disclosed, or represents
that its use would not infringe privately owned rights.  Reference herein
to any specific commercial products process, or service by its trade name,
trademark, manufacturer, or otherwise, does not necessarily constitute or
imply its endorsement, recommendation, or favoring by the United States
Government or any agency thereof, or The Regents of the University of
California.  The views and opinions of authors expressed herein do not
necessarily state or reflect those of the United States Government or any
agency thereof of The Regents of the University of California and shall
not be used for advertising or product endorsement purposes.
\end{quotation}
\end{scriptsize}

\vskip 2in

\begin{center}
\begin{small}
{\it Lawrence Berkeley Laboratory is an equal opportunity employer.}
\end{small}
\end{center}

\newpage
\renewcommand{\thepage}{\arabic{page}}
\setcounter{page}{1}

\section{Introduction}

\hspace{0.8cm} The standard model, while extremely successful, has 18 free
parameters, 13 of which are in the flavor sector. In seeking a more fundamental
theory we can be guided by the requirement that at least some of these
parameters should  be predicted. Symmetries provide essentially the only tool
which is sufficiently developed to yield such predictions: imposing extra
symmetries on a theory leads to a reduction in the number of free parameters.
Such symmetries have been studied in the flavor sector for over 20 years.
Experiment has recently provided a hint as to which symmetries should be
imposed: the only one of the 18 parameters of the standard model which has been
successfully predicted to a high level of significance is the weak mixing angle
[1]. This suggests that we should pursue theories which have both supersymmetry
and grand unified symmetry.

Mass relations from grand unified theories can typically only be checked at the
20-30\% level. This is because many of the masses and mixing angles are not
precisely known, and because the predictions depend on the strong gauge
coupling constant. Hence if predictions from the flavor sector are to be highly
significant, there should be as many of them as possible. This has been a
guiding principle behind several recent works on the charged fermion mass
sector [2,3,4]. The first of these was based on the Georgi-Jarlskog texture
[5], which is a highly successful ansatz for the form of the Yukawa matrices at
the GUT scale involving six independent operators. Including one physical phase
and the ratio of Higgs vacuum expectation values, tan $\beta$, this scheme
describes the 13 observable flavor parameters in terms of 8 free parameters,
thereby predicting 5 of the standard model flavor parameters.

A more recent analysis [4] is much more ambitious. A general operator analysis
is done for the flavor sector of $SO(10)$ theories [6], and several models are
found where just the minimum possible number of operators, 4, successfully
describe flavor physics. This search for a maximally predictive flavor sector
is based on the observation that a set of spontaneously broken family
symmetries generally leads to a hierarchy of operators such that the dominant
contribution to fermion masses arises from just a few such operators[3,5,7].
This reduces the number of free parameters by two compared with the
Georgi-Jarlskog case, and hence predicts 7 of the 13 standard model flavor
parameters. While there is no guarantee that these minimal $SO(10)$ flavor
models are correct, they do merit attention: it is surprising that such
economic descriptions of fermion masses exist, and they have already reached a
level of some significance by virtue of having 7 predictions in agreement with
present measurements. They will be further tested as the flavor parameters
become better measured.

How should these theories be extended to include predictions for neutrino
masses and mixings? It might be thought that no further assumptions are
necessary: since the families fall into 16-dimensional representations of
$SO(10), 16_i \ i = 1,2,3$, and since $16_i \supset u_i, d_i, e_i, \nu_i$ it
could be expected that the operators which generate the Yukawa matrices \bfu,
\bfd, and \bfe for charged fermions will automatically generate the Yukawa
matrices for neutrinos. Unfortunately this is not the case. The neutrino masses
involve both the Dirac Yukawa matrix, $\boldl$, which couples doublet to
singlet neutrinos, and also the Majorana Yukawa matrix, $\boldL$, which
involves only singlet neutrinos. Given a grand unified model which specifies
\bfu, \bfd, and \bfe there are always further assumptions which must be made
for $\boldl$, and especially $\boldL$, to be specified[8].  Indeed, the more
neutrino mass predictions one wants to make the more assumptions must typically
be made.

For example, the Georgi-Jarkskog ansatz can be studied within particular
$SO(10) $ schemes for neutrino mass predictions. If a broad class of models is
studied (as was done by Harvey, Ramond, and Reiss in [5])  few if any precise
numerical predictions can be made. However, it is possible to narrow one's
focus to the case that the necessary 6 $SO(10)$ invariant operators have a very
special form, in which case the maximal number of predictions can be made: both
mass ratios and all mixing angles and phases of the leptonic Kobayashi-Maskawa
matrix [9]. In the case of charged fermion masses it is worth seeking maximally
predictive theories, because they can immediately be tested by their
predictions. This is not the case for models for neutrino masses. While we have
an existence proof of a maximally predictive model [9] no aspect of the
neutrino sector has yet been tested and it may be that maximally predictive
theories of this sort will fall victim to the large number of assumptions
inherent in their construction. However, some principle is required in guiding
a search for grand unified models for neutrino mass prediction. What should
replace the idea of maximal predictivity based on spontaneously broken family
symmetries that was used in the charged fermion mass sector [4]?

In this paper we enumerate a set of minimal assumptions which allows the
prediction of just one quantity, the $\nu_\mu$ to $\nu_\tau$ mixing angle
$\thetut \equiv |V_{\nu_{\mu}\tau} | $ and give the possible predictions for
this angle that result from $SO(10)$ grand unification. The reason for
concentrating on this angle is that we believe that if the grand unified
framework is correct, then this is the parameter of the neutrino sector with
the best prospect of being measured over the coming years. A prediction for
$\thetut$ takes the form:

$$
\thetut = \kappa \abs{ \vcb }\eta\eqno(1.1)
$$
where $\kappa$ is a GUT generalized Clebsch-Gordan group theory factor expected
to have a simple numerical value (eg. 1, 3, 1/3, 2/3, ...) and $\eta$ is a
renormalization group correction which can be computed and is expected to be
close to unity. The present oscillation limits correspond to $\thetut < 0.032$,
and the CHORUS [10] and NOMAD [11] experiments will reduce this to 0.01. Hence
current experiments are precisely at the right sensitivity to check relations
of the form (1.1), assuming $\Delta m^2$   is large enough.

In the grand unified framework the heaviest neutrino is $\nu_\tau$ and hence
the largest $\Delta m^2$ will be for $\nu_e \nu_\tau$ and $\nu_\mu \nu_\tau$
oscillations. Since one typically finds $\theta_{e\tau}$ to be of order
$\abs{V_{ub}}$, the amplitude for these oscillations are prohibitively small.
For CHORUS and NOMAD to see a positive signal, $m_{\nu_\tau}$ would have to be
at least 1 eV. Is this likely to be true? The simplest estimate for this mass
is $m_{\nu_\tau} \approx v^2/V$ where $v$ is the electroweak vev, and $V$ is
the scale of $B-L$ breaking, presumably the GUT scale. This produces the very
disappointing expectation that $m_{\nu_\tau} \approx 10^{-2} eV$. In the models
we discuss below there is a reason why this estimate should be enhanced by
about a factor of 30. Furthermore, the overall magnitude of the neutrino masses
cannot be accurately predicted, the above is simply an order of magnitude
guess. This is to be contrasted with relations of the form of (1.1) which can
give precise predictions at the 20\% level of accuracy. While it is hard to
argue theoretically that $\Delta m^2$ must be large enough for laboratory
observations of $\nu_\mu\nu_\tau$ oscillations, the situation for
$\nu_e\nu_\mu$ oscillations is only worse. Furthermore, there are hints from
both the solar neutrino problem and from dark matter that $\Delta m^2 \approx
100$ eV$^2$.

In this paper we do not consider predictions for other quantities in the
neutrino sector. The reason for this restricted view is the hope that
predictions of the form of (1.1) can be obtained with relatively mild
assumptions. We will find that this hope is only partly borne out. Predictions
can be made based on studies of just two $SO(10)$ invariant operators: the ones
contributing to the 23 and 33 entries of the Yukawa matrices. One need not
consider in any detailed way the operators for the lighter generations.
Furthermore, the operators for the heavier generations are the ones with the
simpler structure. Nevertheless, the list of assumptions which we are forced to
make is still uncomfortably long.

The light neutrino mass matrix is given by

$$
{\bf m_\nu} = {v^2\over V} \boldl{\boldL}^{-1} \boldl^T
\eqno(1.2)
$$
where $v$ = 247 GeV is the electroweak vev, and $V$ is the $B-L$ breaking vev.
It is because $V$ is unknown that the scale of the neutrino masses cannot be
predicted. Once an $SO(10)$ invariant model for \bfu, \bfd, and \bfe has been
written down, the matrix $\boldl$ will follow immediately. At the GUT scale the
contributions to $\lambda_{ij}$ will be Clebschs times the $U_{ij}$
contributions. Hence the difficulty in making predictions for neutrino mass
ratios and mixing angles lies not with $\boldl$, but with $\boldL$. Suppose the
two Higgs doublets of the low energy theory, taken to be the minimal low energy
supersymmetric model, originate from some GUT representations $\{\phi\}$. The
crucial question is whether the singlet field which acquires vev $V$ lies
wholly within this same set $\{\phi\}$. If not then there will be entries of
$\boldL$ which are completely unrelated to those of \bfu, \bfd, \bfe. To avoid
this difficulty we assume that the form of $\boldL$ is such that $\thetut$
depends only on the entries of $\boldl$ and not on those of $\boldL$. While
this is a strong assumption, it nevertheless includes a very large class of
models. It has the very obvious advantage that provided $\boldL$ possesses
certain features which puts it in this class of models, we do not need to know
anything else about $\boldL$ in order to make our prediction.

To convince the reader that there are many models where $\thetut$ does not
depend on the $\Lambda_{ij}$ entries, consider the following. It is
phenomenologically well motivated to have $U_{11} = U_{13} = U_{31} = D_{11}
=D_{13} = D_{31} =0$ [12]. Essentially all predictive ansatzes have this form,
because it results, when $U_{12} = U_{21}$ and $D_{12} = D_{21}$, in the
successful relations $V_{ub}/\vcb = \sqrt{ m_u/m_c}$ and $V_{td}/V_{ts} =
\sqrt{m_d/m_s}$. The $SO(10)$ theory will then give $\lambda_{11} =
\lambda_{13} = \lambda_{31} =0$.

Hence we take

\vskip 9pt
$$
\boldl = \pmatrix{ 0&d&0\cr
                    d'&c&b\cr
                    0&b'&a\cr}\eqno(1.3)
$$
and a completely general $\boldL$:
$$
\boldL = \pmatrix{ F&D&E\cr
                    D&C&B\cr
                    E&B&A\cr}\eqno(1.4)
$$
with
$$
{\boldL}^{-1} = {1\over det \ {\boldL}}
             \pmatrix{ \widetilde{F}&\widetilde{D}&\widetilde{E}\cr
                       \widetilde{D}&\widetilde{C}&\widetilde{B}\cr
                       \widetilde{E}&\widetilde{B}&\widetilde{A}\cr}\eqno(1.5)
$$
where $\widetilde{A} =  CF - D^2, \widetilde {B} = ED - BF,
\widetilde{C} = AF - E^2,
\widetilde{D} = BE - AD, \widetilde{E} = BD - CE$ and $\widetilde{F} = AC-B^2$.

One finds the contributions to $\thetut$ from diagonalization of {\bf $m_\nu$}
of equation (2) to be
$$
\theta^{(\nu)}_{\mu\tau} = {b'd'\widetilde{D} + ad' \widetilde{E} + b'c
\widetilde{C} + (ac + bb') \widetilde{B} + ab\widetilde{A}\over
b'^2\widetilde{C} + 2ab'\widetilde{B} + a^2\widetilde{A} }\eqno(1.6)
$$
If all terms in numerator and denominator are comparable, there is little hope
for a prediction for $\thetut$. However, if the $\widetilde{A}$ terms dominate
then $\thetut = b/a$ depends only on $\boldl$ not on $\boldL$. If
$\widetilde{B}$ dominates the $\thetut= {b\over 2a} + {c\over 2b'}$, while if
$\widetilde{C}$ dominates $\thetut = c/b'$. In this paper we will assume that
either $\widetilde{A}$ or $\widetilde{B}$ dominates, and that the resulting
prediction is either $b/a$ or $b/2a$. (i.e. in the latter case we assume $c/b'
\ll b/a$). The resulting predictions depend only on the 23 and 33 entries of
$\boldl$, which come from simple operators chosen to yield correct heavy
generation masses and $\vcb$. These are the cases which  lead to predictions of
the form of equation (1.1).

There are several ways that $\widetilde{A}$ or $\widetilde{B}$ dominance can
occur. For example, suppose that $\boldL$ has the same pattern of zeros as the
other matrices so that $\Lambda_{11} = F =0$ and $\Lambda_{13} = E = 0$. In
this case $\widetilde{B} = \widetilde{C} =0$ so that (1.6) reduces to

$$
\theta^{(\nu)}_{\mu\tau} = {b'd' \widetilde{D} + ad'\widetilde{E} + ab
\widetilde{A}\over a^2\widetilde{A}} \eqno(1.7)
$$
with $\widetilde{A} = -D^2, \widetilde{D} = -AD$ and $\widetilde{E} = BD$. We
will argue below that the simplest operator for the 33 entry which gives masses
to the heaviest generation leads to $A =0$ so $\widetilde{D}$ can be dropped.
In the numerator of (1.7) there is a simple competition between two terms and
presumably in a reasonable fraction of theories the $\widetilde{A}$ term
dominates.

We are now ready to list our assumptions

(1) We study a supersymmetric grand unified $SO(10)$ theory, broken at the
scale of grand unification to the MSSM.

(2) We assume $\thetut$ does not involve the elements of the Majorana Yukawa
matrix $\boldL$, and that

$$
\theta^{(\nu)}_{\mu\tau} = {\lambda_{23}\over \lambda_{33}} \, \ \hbox{or } \
\half {\lambda_{23}\over \lambda_{33}}\eqno(1.8)
$$

(3) The 33 entries of \bfu,\bfd,\bfe, and $\boldl$ matrices arise from a single
operator. The unique possibility which gives a successful top mass prediction
is $O_1 = 16_3 \ 10\  16_3$ [13].

(4) The dominant contribution to the 23 entries of the \bfu,\bfd,\bfe, and
$\boldl$ matrices arises from a single operator $O_2$.

(5) We also demand that either $D_{23} = E_{23} = 0$, or that $O_2$ contains
the same 10 as $O_1$. The reason for this assumption will emerge below.

\section{GUT Scale Predictions}

\hspace{0.8cm} Let us now use these assumptions to make predictions for
$\thetut$. The \bfu,\bfd,\bfe, and $\boldl$ matrices of the two heaviest
generations take the following forms:
\begin{eqnarray*}
  \bfu = \pmatrix{U_{22}&B_U\cr
                     U_{32}&A_U\cr}&
  \bfd = \pmatrix{D_{22}&B_D\cr
                     D_{32}&A_D\cr}\\
  \boldl = \pmatrix{\lambda_{22}&\chi B_U\cr
                           \lambda_{32}&A_U\cr}&
  \bfe = \pmatrix{E_{22}&\epsilon B_D\cr
                     E_{32}&A_D\cr}
\eqno{(2.1)}
\end{eqnarray*}
where $\epsilon$ and $\chi$ are group Clebschs and where all matrix elements
are a priori complex before phase redefinitions on the fields.  $A_U$ and $A_D$
are not equal in general because the two light $SU(2)$ doublets do not
necessarily lie entirely in the 10 which generates the 33 entries. The mass
hierarchy in different generations requires that the 33 entries be much larger
than the other matrix elements, in which case  $|\vcb|$ and $\thetut$ at GUT
scale are given by

$$
\abs{\vcb^{0}} = \abs{{B_U\over A_U}-{B_D\over A_D}}\eqno{(2.2)}
$$

$$
\thetut^{0} = \abs{{\chi B_U \over A_U}-{\epsilon B_D \over A_D}}
\hskip .1in  \mbox{or} \hskip .1in
\abs{{\chi B_U \over 2 A_U}-{\epsilon B_D \over A_D}}\eqno{(2.3)}
$$
where the superscript $``0"$ represents the quantity at the GUT scale . In
general  $B_U \over A_U$ and $B_D \over A_D$ are unrelated and  cannot be made
real simultaneously by  redefining the relative phases of the quark fields. The
unknown relative phase and magnitude between $B_U \over A_U$ and $B_D \over
A_D$ prevent us   from making any definite predictions. Therefore, to get the
simple relation (1.1) requires  that either $B_D=0$ or  that $O_2$ contains the
same 10 as $O_1$ in which case $B_D \over A_D$ and $B_U \over A_U$ have the
same phase and are related by a  group  Clebsch factor $\xi$ (this is the
explanation for assumption \#5 above).  For either case we can write:

$$
{B_D \over A_D} = \xi {B_U \over A_U} \hskip 0.5in (\mbox{where} \hskip .1in
 \xi = 0 \hskip .1in
\mbox{if} \hskip .1in B_D=0)\eqno{(2.4)}
$$
Then
$$
  \abs{\vcb^0} = \abs{(1-\xi){B_U \over A_U}}\eqno{(2.5)}
$$
\begin{eqnarray*}
  \thetut^0 = \abs{(\chi-\epsilon\xi){B_U \over A_U}}
  \hskip .1in \mbox{or} \hskip .1in
  \abs{({\chi \over 2}-\epsilon\xi){B_U \over A_U}}\eqno{(2.6)}\\
\end{eqnarray*}

 We then find the following $\kappa$ factors in eq. (1.1):

\begin{eqnarray*}
  \Rightarrow\ \kappa = \abs{{\chi-\epsilon\xi \over 1-\xi}} (\equiv \kappa_1)
  \hskip .1in \mbox{or} \hskip .1in
  \abs{{{\chi \over 2}-\epsilon\xi \over 1-\xi}} (\equiv \kappa_2)\eqno{(2.7)}
\end{eqnarray*}

In what follows we study extensively the possible $O_2$'s and then calculate
the various $\kappa$ factors that result from them.

\subsection{Operators Contributing to the Two-Three Yukawa Matrix Elements}

\hspace{0.8cm} For $O_2$ being a dimension-4 operator, there are only 3
possibilities, $16_2 \ 10\ 16_3$, $16_2\ 120\ 16_3$, and $16_2\ \overline{126}
\ 16_3$. However, $\vcb\approx {1 \over 20}$ hints that $O_2$ may be a
higher-dimension operator which is suppressed by some powers of masses of order
$M_P$ (for examples, see Table 3).  For simplicity we will restrict ourselves
to dimension-5 operators.  Also,  if the 23 entry is generated by a dimension-6
or higher operator, then the expansion in powers of ${v \over M}$ is $\gsim {1
\over 4} \rightarrow {1 \over 5}$ which is becoming uncomfortably large.

A dimension-5 operator has the form

\ba
\eqno{(2.8)}
{1 \over \mpp} \sixttw \phione \phitwo  \sixtth
\ea
To make an $SO(10)$ invariant the product of the $\phione\times \phitwo$ must
contain a 10, 120, or $\overline{126}$ of $SO(10)$ since $16 \times 16 = 10 +
120 + 126$ (the 10 and 120 of SO(10) are self-conjugate). For representations
of dimension $\leq 126$, the possible combinations of $\phione\phitwo$ are:

\ba
  45\ \times \ 10 \ & \supset &\ 10,\ 120,\\
\eqno{(2.9a)}
  45\ \times \ 120 \ & \supset &\ 10,\ 120,\ \overline{126},\\
  45\ \times \ \overline{126} \ & \supset &\ 120,\ \overline{126},\\
\ea
\ba
  54\ \times \ 10 \ & \supset &\ 10,\\
\eqno{(2.9b)}
  54\ \times \ 120 \ & \supset &\ 120,\\
  54\ \times \ 126 \ & \supset &\ \overline{126},\\
\ea
\ba
  16\ \times \ 16 \ & \supset &\ 10,\ 120,\\
\eqno{(2.9c)}
  \overline{16}\ \times \ \overline{16} \ &\supset &\ 10,\ 120,\
\overline{126}.
\ea

$45\ \times\ 126$ also contains 120, but this does not couple to $16\ \times\
 16$.

For operators (2.9a), we consider the case in which the 45 vev lies in a
definite direction in the $SO(10)$ group space: in one of the  hypercharge $Y$,
$B-L$, $T_{3R}$, or $X$ directions, where $X$ preserves the $SU(5)$ subgroup.
In fact, $SO(10)$ may be broken to $SU(5)$ by a vev of $45_X$ at a larger scale
than $SU(5)$ is broken. This means that the mass scale appearing in the
denominator in (2.8) can be $\VEV{45_X}$ (on the order of the GUT scale)  as
well as masses of order $M_P$. When a 45 vev acts on a fermion in a 16, it
gives a numerical Clebsch which is the charge of the fermion under the
particular group generator corresponding to the direction of this vev.
Therefore the 23 entries of the fermion mass matrices are just related by these
Clebschs. These Clebschs are listed in Table 1.

\begin{table}[htbp]
$$
  \begin{array}{|c|r|r|r|r|}  \hline
       &  X         &  Y         & B-L            &  T_{3R}           \\ \hline
  q    &  1         &  1         &  1            &  0                \\ \hline
  u^c  &  1         & -4         & -1            & -1                \\ \hline
  d^c  & -3         &  2         & -1            &  1                \\ \hline
  l    & -3         & -3         & -3            &  0                \\ \hline
  e^c  &  1         &  6         &  3            &  1                \\ \hline
  N^c  &  5         &  0         &  3            & -1                \\ \hline
  \end{array}
$$
\caption{The charges (normalized to integers) of fermion fields under the group
         generators $X$, $Y$, $B-L$, and $T_{3R}$.}
\end{table}

For operators (2.9b), the product of a 54 and another representation  contains
only one of the 10, 120 and $\overline{126}$, so the predictions are the same
as those of the dimension-4 operators and need not be discussed separately.

The $16^2$ and $\overline{16}^2$ of (2.9c) are more complicated. Under the
$SU(5)$ decomposition, $16=1+\overline{5}+10$ and $\overline{16}=1+5+
\overline{10}$. We can see that $16^2$ only contributes to the masses of down
quarks and charged leptons and $\overline{16}^2$ only contributes to the masses
of up quarks and neutrinos (because the former arises from the $SU(5)$ operator
$10 \times \fbar \times \fbar$ and the latter from $10\times 5 \times 10$).
Therefore, if $O_2 = {B \over M}\ 16_2 \ \overline{16}^2 \ 16_3$, it is natural
to have $D_{23}=E_{23}=0$. Now the  vev of the 5 of $SU(5)$ breaks $SU(2)_L$ so
it must be light. Only the singlet of $SU(5)$ could develop a vev at the order
of the GUT scale.  This operator is more complicated because in addition to the
Dirac masses, it also gives Majorana masses to both left handed and right
handed neutrinos. In this case, the light neutrino mass matrix receives an
additional contribution beyond that of eq. (1.2) and we discuss this below.

 We are now ready to list the possible operators and the resulting Clebsch
coefficients.

\subsection{The Clebsch Coefficients $\kapo$ and $\kapt$}

1. Dimension-4 operators:

(1) $O_2 = 16_2\ 10\ 16_3$, where the 10 is the same 10 as in $O_1$. In this
case we have ${B_U \over A_U}={B_D \over A_D}\ \Rightarrow\ \vcb=0$, which
conflicts with experiment, so it's excluded.

(2) $O_2 = 16_2\ 10^{\prime}\ 16_3$, where the $10'$ is different from the 10
in $O_1$ and has a vev contributing to \bfu and $\boldl$ only. In this case
$\xi =0,\ \chi =1$ and therefore $\kappa_1 =1,\ \kappa_2 ={1 \over 2}$.

(3) $O_2 = 16_2\ \overline{126}\ 16_3$. The $\overline{126}$ contributes to the
up quark and the Dirac neutrino mass matrix elements with the ratio $U_{23} :
\lambda_{23} = 1 : -3$. One obtains $\xi =0,\ \chi =-3$ and $\kappa_1 =3,\
\kappa_2 ={3 \over 2}$.

(4) $O_2 = 16_2\ 120\ 16_3$. 120 contains 2 independent vevs which contribute
to the masses of up quarks and neutrinos. To be able to get predictions we have
to assume that the vev lies in some particular direction. In the $SU(5) \times
U(1)$ decomposition the 2 vevs lie in the 5 and 45 directions of $SU(5)$ and
contribute to the neutrino and up quark masses separately, so $\lambda_{23}$
and $U_{23}$ are not related. In the $SU(4) \times SU(2) \times SU(2)$
decomposition, the 2 vevs lie in the (2,2,1) and (2,2,15) directions which give
the ratios $U_{23}:\lambda_{23}=1:1$ and $U_{23}:\lambda_{23}=1:-3$
respectively. The former is identical to the case of $10'$ and the latter is
identical to the case of $\overline{126}$.

The Clebschs for dimension-4 operators are summarized in Table 2.
\begin{table}[htbp]
$$
  \begin{array}{|c|r|r|r|r|}  \hline
  Operator     &  \xi         &  \chi         & \kapo  & \kapt \\ \hline
  \sixttw\ 10\ \sixtth & \multicolumn{4}{|c|}{\vcb=0}             \\ \hline
  \sixttw\ 10'\ \sixtth & 0 & 1 & 1 & {1 \over 2}                \\ \hline
  \sixttw\ \otsbar\ \sixtth & 0 & -3 & 3 & {3 \over 2}                \\ \hline
  \sixttw\ 120\ \sixtth & \multicolumn{4}{|c|}{\mbox{Same as $10'$ and
           $\otsbar$}}
                           \\ \hline
  \end{array}
$$
\caption{The Clebschs for all dimension-4 operators.}
\end{table}

2. Dimension-5 operators, the 2 scalars can be:

(1) $10 \times 45$, where 10 is the same 10 in $O_1$. The Clebschs for
different operators are listed in Table 3.
\begin{table}[htbp]
$$
  \begin{array}{|l|r|r|r|r|r|}   \hline
  \multicolumn{1}{|c|}{\mbox{Operator}}
   & \xi & \chi & \epsilon & \kappa_1 & \kappa_2 \\ \hline
  16_2\ 10\ {45_X \over M}\ 16_3 & -3 & 5 & -{1 \over 3} & 1 & {3 \over 8}
  \\ \hline
  16_2\ 10\ {45_Y \over M}\ 16_3 & -{1 \over 2} & 0 & 3 & 1 & 1 \\ \hline
  16_2\ 10\ {45_{B-L} \over M}\ 16_3 & 1 & -3 & -3 &
   \multicolumn{2}{c|}{\vcb=0} \\ \hline
  16_2\ 10\ {45_{T_{3R}} \over M}\ 16_3 & -1 & 1 & 1 & 1 & {3 \over 4} \\
\hline
  16_2\ {45_{X,Y,B-L,T_{3R}} \over M}\ 10\ 16_3 & \multicolumn{5}{c|}{\vcb=0}
  \\ \hline
  16_2\ 10\ {45_Y \over \ffxv}\ 16_3 & {1 \over 6} & 0 & -9 & {9 \over 5} &
  {9 \over 5} \\ \hline
  16_2\ 10\ {45_{B-L} \over \ffxv}\ 16_3 & -{1 \over 3} & -{3 \over 5} & 9 &
  {9 \over 5} & {81 \over 40} \\ \hline
  16_2\ 10\ {45_{T_{3R}} \over \ffxv}\ 16_3 & {1 \over 3} & {1 \over 5} & -3 &
  {9 \over 5} & {33 \over 20} \\  \hline
  16_2\ {45_{X,Y,B-L,T_{3R}} \over \ffxv}\ 10\ 16_3 &
\multicolumn{5}{c|}{\vcb=0}
  \\ \hline
  \end{array}
$$
\caption{The Clebschs for the operators containing scalars $10 \times 45$.}
\end{table}

The following models all have $\xi =0$.

(2) $10' \times 45 $ (different 10 than in $O_1$).
 The results are listed in Table 3.

\begin{table}[htbp]
$$
  \begin{array}{|l|r|r|r|}    \hline
  \multicolumn{1}{|c|}{\mbox{Operator}}
  & \chi & \kappa_1 & \kappa_2 \\ \hline
  16_2\ 10'\ {45_X \over M}\ 16_3 & 5 & 5 & {5 \over 2} \\ \hline
  16_2\ 10'\ {45_Y \over M}\ 16_3 & 0 & 0 & 0 \\ \hline
  16_2\ 10'\ {45_{B-L} \over M}\ 16_3 & -3 & 3 & {3 \over 2} \\ \hline
  16_2\ 10'\ {45_{T_{3R}} \over M}\ 16_3 & 1 & 1 & {1 \over 2} \\ \hline
  16_2\ {45_X \over M}\ 10'\ 16_3 & -3 & 3 & {3 \over 2} \\ \hline
  16_2\ {45_Y \over M}\ 10'\ 16_3 & -3 & 3 & {3 \over 2} \\ \hline
  16_2\ {45_{B-L} \over M}\ 10'\ 16_3 & -3 & 3 & {3 \over 2} \\ \hline
  16_2\ {45_{T_{3R}} \over M}\ 10'\ 16_3 & \multicolumn{3}{c|}
  {\vcb=0}  \\ \hline
  16_2\ 10'\ {45_Y \over \ffxv}\ 16_3 & 0 & 0 & 0 \\ \hline
  16_2\ 10'\ {45_{B-L} \over \ffxv}\ 16_3 & -{3 \over 5} & {3 \over 5} &
  {3 \over 10} \\ \hline
  16_2\ 10'\ {45_{T_{3R}} \over \ffxv}\ 16_3 & {1 \over 5} &
  {1 \over 5} & {1 \over 10} \\ \hline
  16_2\ {45_Y \over \ffxv}\ 10'\ 16_3 & 1 & 1 & {1 \over 2} \\ \hline
  16_2\ {45_{B-L} \over \ffxv}\ 10'\ 16_3 & 1 & 1 & {1 \over 2} \\ \hline
  16_2\ \ {45_{T_{3R}} \over \ffxv}\ 10'\ 16_3 &
  \multicolumn{3}{c|}{\vcb=0} \\ \hline
  \end{array}
$$
\caption{The Clebschs for the operators containing scalars $10' \times {45}$.}
\end{table}

(3) $\overline{126} \times 45$. The results are listed in Table 4.

\begin{table}[htbp]
$$
  \begin{array}{|l|r|r|r|}      \hline
  \multicolumn{1}{|c|}{\mbox{Operator}}
  & \chi & \kappa_1 & \kappa_2 \\ \hline
  16_2\ \overline{126}\ {45_X \over M}\ 16_3 & -15 & 15 & {15 \over 2} \\
\hline
  16_2\ \overline{126}\ {45_Y \over M}\ 16_3 & 0 & 0 & 0 \\ \hline
  16_2\ \overline{126}\ {45_{B-L} \over M}\ 16_3 & 9 & 9 & {9 \over 2} \\
\hline
  16_2\ \overline{126}\ {45_{T_{3R}} \over M}\ 16_3 & -3 & 3 & {3 \over 2}
  \\ \hline
  16_2\ {45_X \over M}\ \overline{126}\ 16_3 & 9 & 9 & {9 \over 2} \\ \hline
  16_2\ {45_Y \over M}\ \overline{126}\ 16_3 & 9 & 9 & {9 \over 2} \\ \hline
  16_2\ {45_{B-L} \over M}\ \overline{126}\ 16_3 & 9 & 9 & {9 \over 2} \\
\hline
  16_2\ {45_{T_{3R}} \over M}\ \overline{126}\ 16_3 &
  \multicolumn{3}{c|}{\vcb=0} \\ \hline
  16_2\ \overline{126}\ {45_Y \over \ffxv}\ 16_3 & 0 & 0 & 0 \\ \hline
  16_2\ \overline{126}\ {45_{B-L} \over \ffxv}\ 16_3 & {9 \over 5} &
  {9 \over 5} & {9 \over 10} \\ \hline
  16_2\ \overline{126}\ {45_{T_{3R}} \over \ffxv}\ 16_3 & -{3 \over 5} &
  {3 \over 5} & {3 \over 10} \\ \hline
  16_2\ {45_Y \over \ffxv}\ \overline{126}\ 16_3 & -3 & 3 & {3 \over 2} \\
\hline
  16_2\ {45_{B-L} \over \ffxv}\ \overline{126}\ 16_3 & -3 & 3 & {3 \over 2}
  \\ \hline
  16_2\ {45_{T_{3R}} \over \ffxv}\ \overline{126}\ 16_3 &
  \multicolumn{3}{c|}{\vcb=0}  \\ \hline
  \end{array}
$$
\caption{The Clebschs for the operators containing scalars $\overline{126}
\times 45$.}
\end{table}

(4) $120 \times 45$. Similar to the $16_2\ 120\ 16_3$, either no prediction
can be made or it gives the same results as those of the $10' \times 45$
and $\overline{126} \times 45$ cases.

(5) $\overline{16}^2$. Predictions can be obtained if the 2 $\overline{16}$'s
are the same otherwise there will be too many arbitrary parameters to be fixed.
 Under $SU(5)$, $\overline{16}=1+5+\overline{10}$. The 1 gets a vev $\VEV{1}$
on the order of the GUT scale and the 5 gets a vev $\VEV{5}$ on the order of
the weak scale. While the product $\vev{1}\vev{5}$ contributes to the Dirac
masses of up quarks and neutrinos, the $\vev{1}\vev{1}$ and $\vev{5}\vev{5}$
contribute only to the Majorana masses of the right handed and the left handed
neutrinos respectively. When there is a direct contribution to the light
neutrino mass matrix, equation (2) must be modified to

$$
{\bf m_\nu} = \tilde{v} {\boldnu} - {v^2\over V}\boldl{\boldL}^{-1}
\boldl^T \eqno(16)
$$
where ${\boldnu}$ is the  Yukawa matrix of the left handed neutrinos and
$\tilde{v}$ has  mass dimension unity. Here we choose $\tilde{v}={v^2\over V}$
so that ${\boldnu}$ can be compared with $\boldl$ and $\boldL$. If this
operator is the dominant source of the Majorana masses of the neutrinos
($\widetilde{B}$ dominates in equation 6), $({\bf m_\nu})_{23}$ receives
comparable contributions from both terms as we can see the first term $\sim
(\vev{5})^2$ and the second term $\sim {(\vev{1} \vev{5})^2 \over
(\vev{1})^2}$. From the group theory we know $16 \times \overline{16} \supset
1, 45, 210, \hskip .1in \overline{16}^2 \supset 10, \overline{126}$ (no 120
because the 2 $\overline{16}$'s are the same and 120 is antisymmetric). There
are several ways to combine these fields into $SO(10)$ invariants.

(i) $(16_2\ \overline{16})_1\ (\overline{16}\ 16_3)_1$: This operator does not
contribute to $U_{23}$ and is thus not interesting.

(ii) $(16_2\ \overline{16})_{45}\ (\overline{16}\ 16_3)_{45}$: This gives the
ratio $U_{23}:\lambda_{23}:\Lambda_{23}:\nu_{23} = 8:3:-5:-5 \Rightarrow \chi =
{3 \over 8}$. The ratio of the two terms in $({\bf m_\nu})_{23}$ is $-5:{3^2
\over -5}$ so $\kappa_2$ has to be modified by a factor $\abs{{5^2 \over
3^2}-1} = {16 \over 9}$. In this case $\kappa_2 = \abs{{16 \over 9} \times
{\chi \over 2}} = {1 \over 3}$.

(iii) $(16_2\ \overline{16})_{210}\ (\overline{16}\ 16_3)_{210}$: This gives
the ratio $U_{23}:\lambda_{23}:\Lambda_{23}:\nu_{23} =4:9:5:5 \Rightarrow \chi
= {9 \over 4}$. Similarly $\kappa_2$ is modified by a factor $\abs{{5^2 \over
9^2}-1} = {56 \over 81}$. In this case $\kappa_2 = \abs{{56 \over 81} \times
{\chi \over 2}} = {7 \over 9}$.

(iv) $(16_2\ 16_3)_{10}\ (\overline{16}\ \overline{16})_{10}$: This operator
does not contribute to the Majorana masses so the predictions are the same as
those of $16_2\ 10'\ 16_3$, \ie , $\kappa_2 = {1 \over 2}$.

(v) $(16_2\ 16_3)_{126}\ (\overline{16}\ \overline{16})_{\overline{126}}$:
This gives the ratio $U_{23}:\lambda_{23}:\Lambda_{23}:\nu_{23} = 1:-3:-8:-8
\Rightarrow \chi =-3$. $\kappa_2$ is modified by a factor $\abs{{8^2 \over
3^2}-1}  = {55 \over 9}$. Therefore we have $\kappa_2 = \abs{{55 \over 9}
\times {\chi \over 2}} = {55 \over 6}$.

The $\kappa_1$ predictions result from the condition that $\widetilde{A}$
dominates ($AD^2 \gg B^2 F$, ${\boldL}^{-1}$ is dominated by $\Lambda_{33}$) in
equation (1.6), so there is no simple relation between the two terms in $({\bf
m_{\nu}})_{23}$. Predictions can be obtained only when the first term is
negligible, or in terms of $\boldL$ matrix elements, $A \ll B$. In this case,
$\kappa_1 = {3 \over 8}, {9 \over 4}, 1, 3$ for the operators (ii), (iii),
(iv), and (v) respectively. The results are summarized in Table 6.

\begin{table}[htbp]
$$
  \begin{array}{|l|r|r|r|}      \hline
  \multicolumn{1}{|c|}{\mbox{Operator}}
  & \chi & \kappa_1 & \kappa_2 \\ \hline
(16_2\ \overline{16})_1\ (\overline{16}\ 16_3)_1 & \multicolumn{3}{|c|}{-}
 \\ \hline
(16_2\ \overline{16})_{45}\ (\overline{16}\ 16_3)_{45} & {3 \over 8} &
 {3 \over 8} & {1 \over 3} \\ \hline
(16_2\ \overline{16})_{210}\ (\overline{16}\ 16_3)_{210} & {9 \over 4} &
 {9 \over 4} & {7 \over 9} \\ \hline
(16_2\ 16_3)_{10}\ (\overline{16}\ \overline{16})_{10} & 1 & 1 & {1 \over 2}
 \\ \hline
(16_2\ 16_3)_{126}\ (\overline{16}\ \overline{16})_{\otsbar} & -3 & 3
 & {55 \over 6} \\ \hline
  \end{array}
$$
\caption{The Clebschs for the $16^2\ \overline{16}^2$ operators}
\end{table}


\section{The Renormalization Factor $\eta$}

\hspace{0.8cm} In this section we shall discuss the renormalization group (RG)
dependence of the predictions on the mass scale, $t=ln(\mu)$.  As discussed,
the predictions we have found all take the form of equation (1.1) with the
$\kappa$ predictions as shown in the last section. The $\kappa$'s are pure
group theory coefficients and are completely determined by the matrix texture
and mass-producing operators one chooses, and the relations involving these
generalized Clebschs are assumed to hold strictly at the GUT scale.  These
coefficients involve no dynamic information about the RG running down to the
weak scale or the actual experimental inputs, and we have chosen to include all
of this dependence in the $\eta$ factor.   In fact, both $\thetut$ and $\vcb$
renormalize (differently) and the original form of the prediction at the GUT
scale, $\thetuto =\kappa \abs{\vcbo}$, when run down to the electroweak scale
at approximately $\mt$ becomes equation (1.1),  where the unsuperscripted
$\thetut$ and $\vcb$  are taken to be the values at $\mt$ (taken to be on the
order of the SUSY-breaking scale so that our MSSM RGE's are valid down to this
scale).

\subsection{Heavy Yukawa Coupling RGE's}

\hspace{0.8cm} The procedure for obtaining these $\eta$ factors is as follows:
we must first solve the RGE's for the running of the three heaviest generation
Yukawa couplings[14]:

\ba
{d(\htee) \over dt} & = &-\left( {\htee \over \zf} \right)
			\left( 6\htee^2+\hb^2-\gu \right)
\\
\eqno{(3.1)}
{d(\hb) \over dt} & = &-\left( {\hb \over \zf} \right)
			\left( 6\hb^2+\htee^2-\gd \right)
\\
{d(\htau) \over dt} & = &-\left( {\htau \over \zf} \right)
			\left( 3\hb^2+4\htau^2-\gee \right)
\ea
where the $G_{(U,D,E)}$ are the parts of the equation that come from
gauge boson renormalization:

\ba
\gu & = & {13 \over 15}\gon ^2 + 3\gtw ^2+{16 \over 3}\gth ^2
\\
\eqno{(3.2)}
\gd & = &{7 \over 15}\gon ^2 + 3\gtw ^2+{16 \over 3}\gth ^2
\\
\gee & = &{9 \over 5}\gon ^2 + 3\gtw ^2
\ea

These equations must be solved numerically, subject to the boundary conditions
$\hb(\mg)=\htau(\mg)$ (which is not in general equal to  $\htee(\mg)$ since we
have chosen to solve the more general case of $\au \neq \ad$), and ${\hb(\mt)
\over \htau(\mt)} = {\mb(\mt) \over \mtau(\mt)}$.  Here, we must run the
extracted experimental values of $\mb(\mb)$ and $\mtau(\mtau)$ up to the values
at $\mt$, which involves only a gauge boson renormalization over this range:
for $\mb$ we divide through by the factor $\etab$ [2], and we can actually take
$\mtau(\mt) \approx \mtau(\mtau)$ since lepton masses renormalize only by QED
and weak interactions, and thus change very little over this range. The factor
$\etab$ depends on the value of $\alphas (M_Z)$. Since we have not specified
the value of ${\htee(\mg) \over \hb(\mg)}$,  we must fix a value for this in
order to obtain solutions for $\htee,\hb$ and $\htau$.  This is equivalent to
varying the value of $\tb$ we choose, since fixing $\tb$ will fix the value of
$\htau(\mtau)$ (through $\mtau(\mtau) = { {\bf v} \over \sqrt{2} } \htau(\mtau)
\cb$). This then fully determines the solutions for $\htee(\mu), \hb(\mu)$ and
$\htau(\mu)$. Using these solutions we can then determine the definite
integrals of the these heavy Yukawas squared from $\mt$ to $\mg$, which are
factors in the definitions of the $\eta$'s as we shall now see.

\subsection{RGE's for $\vcb$ and $\thetut$}

\hspace{0.8cm} The RGE's for $\vcb$ and $\thetut$ are [15]:

\ba
{d\vcb \over dt} & = & -\left( {\vcb \over \zf} \right)
           \left( \htee^2 + \hb^2 \right)
\\
\eqno{(3.3)}
{d\thetut \over dt} & = & -\left( {\thetut \over \zf} \right)
   \left( \htau^2 \right)
\ea
These equations have the analytic solutions:

\ba
\thetut & = & \thetuto \; \ftaup
\\
\eqno{(3.4)}
\vcb & = & \vcbo\; \ftp \; \fbp
\ea
with

\ba
f_i & = & exp\left(\int^{\mg}_{\mz} h_i^2 dt \right)
\\
\eqno{(3.5)}
\ea
i=t,b, $\tau$.

Thus $\thetuto=\kappa \abs{\vcbo}$ becomes $\thetut  =  \kappa \abs{\vcb} \eta$
with

\ba
\eta & = & {\ftaup \over {\ftp\fbp} }
\\
\eqno{(3.6)}
\ea
The renormalization correction $\eta$ is shown in Figure 1 as a
function of $\tan \beta$ for different $\alphas$ values.

\section{Discussion of Results.}

\hspace{0.8cm} From Figure 1 we can see that the renormalization correction
$\eta$ has strong dependence on $\alphas (M_Z)$ and a somewhat weaker
dependence on $\mb$ and the unknown $\tb$.  This makes our predictions less
definite. With all uncertainties included, $\eta = 0.74 \pm 0.11$ and $\vcb
\eta = 0.033 \pm 0.007$. Therefore the cases of $\kappa =1$ just lie on the
edge of the current experimental limit set by FNAL E531 (as shown in Fig. 2).
The CHORUS and NOMAD experiments will test $\thetut$ within the range
0.01-0.032 which corresponds to $\kappa =0.3$-1. Most of the models we have
discussed have $\kappa > 1$ and therefore large $\Delta m^2$ is already
excluded in these models (even taking uncertainties into account). However,
there are still many models which give $\kappa$ values which are allowed. These
models are listed in Table 7 (except the uninteresting case $\kappa=0$, which
would mimic no $\nu_\mu \nu_\tau$\ oscillations).

\begin{table}[htbp]
$$
  \begin{array}{|l|c|c|}   \hline
  \mbox{Operator} & \kappa_1 & \kappa_2 \\  \hline
  16_2\ 10'\ 16_3 & 1 & {1\over 2} \\ \hline
  16_2\ 10'\ {45_{T_{3R}} \over M}\ 16_3 & 1 & {1\over 2} \\ \hline
  16_2\ {45_Y \over \ffxv}\ 10'\ 16_3 & 1 & {1\over 2} \\ \hline
  16_2\ {45_{B-L} \over \ffxv}\ 10'\ 16_3 & 1 & {1\over 2} \\ \hline
  (16_2\ 16_3)_{10}\ (\overline{16}\ \overline{16})_{10} & 1 & {1\over 2} \\
\hline
  16_2\ 10\ {45_X \over M}\ 16_3 & 1 & {3\over 8} \\ \hline
  16_2\ 10\ {45_{T_{3R}} \over M}\ 16_3 & 1 & {3\over 4} \\ \hline
  16_2\ 10\ {45_Y \over M}\ 16_3 & 1 & 1 \\ \hline
  16_2\ 10'\ {45_{B-L} \over \ffxv}\ 16_3 & {3\over 5} & {3\over 10} \\ \hline
  16_2\ 10'\ {45_{T_{3R}} \over \ffxv}\ 16_3 & {1\over 5} & {1\over 10} \\
\hline
  16_2\ \overline{126}\ {45_{T_{3R}} \over \ffxv}\ 16_3 & {3\over 5} & {3\over
10} \\ \hline
  16_2\ \overline{126}\ {45_{B-L} \over \ffxv}\ 16_3 & - & {9\over 10} \\
\hline
  (16_2\ \overline{16})_{45}\ (\overline{16}\ 16_3)_{45} & {3\over 8} & {1\over
3} \\ \hline
  (16_2\ \overline{16})_{210}\ (\overline{16}\ 16_3)_{210} & - & {7\over 9} \\
\hline
  \end{array}
$$
\caption{The $\kappa$ values for which large $\Delta m^2$
is not excluded.
(Note: The 120 can be substituted for the $10'$ or $\overline{126}$.
See Sec.\ 2.2, Dimension-4 operators, \#4.)}
\end{table}

The preferred $\kappa$ values are 1 and ${1 \over 2}$ as they occur most in
these allowed models. The corresponding $\sin^2 2\thetut$ values are $4\times
10^{-3}$ and $1 \times 10^{-3}$, respectively. We can see from Figure 2 that
they are well within the region which will be probed by the CHORUS and NOMAD
experiments (the NOMAD and CHORUS limits are comparable). In fact, all $\kappa$
values except those given by $16_2\ 10'\ {45_{T_{3R}} \over \ffxv}\ 16_3$ in
Table 7 can be probed by these experiments.

\section{Conclusion}

\hspace{0.8cm} There are two ideas which underlie much of the thinking about
neutrino masses and mixings:
\begin{itemize}
\item Neutrinos are light because of the see-saw mechanism (eq. 1.2), with a
large mass responsible for the Majorana masses.
\item Neutrino mixing is expected to be broadly similar to quark mixing,
because, at some fundamental level, leptons are similar to quarks.
\end{itemize}

These ideas become concrete in the framework of grand unified theories, which
provide both a unification of quarks and leptons and a very large mass scale
for the right handed neutrinos. It is well-known that the idea of grand
unification suggests a hierarchy of neutrino masses: $m_{\nu_e} : m_{\nu_\mu} :
m_{\nu_\tau} \approx m_u : m_c : m_t$ (or perhaps $ m_u^2 : m_c^2 : m_t^2$) and
of neutrino mixing angles: $\theta_{ij} \approx \abs{V_{ij} }$. To what extent
can these suggestions be sharpened into precise numerical predictions for
parameters of the neutrino sector?

In this paper we have shown that a sequence of five assumptions about the grand
 unified theory allows the mixing angle $\thetut$ to be precisely predicted via
eq. (1.1). However there are many Clebschs, $\kappa$, which can appear in such
a prediction, and we have no reason to expect that any one is preferred. Hence
we have searched for all such possible Clebschs. With the exception of just one
case, CHORUS and NOMAD  will discover $\numt$ oscillations, if $\Delta m^2$ is
large enough. However, they will not be able to provide a significant numerical
test of any of our predictions. There are two reasons for this.  The first is
that given the smearing of our predictions due to the experimental
uncertainties of $\alphas$ and $\vcb$, and given the large number of possible
$\kappa$ values, our predictions span the entire region which CHORUS and NOMAD
will probe.  A considerable improvement in this situation can be expected in
the future as the uncertainties, especially on $\vcb$, will be reduced.  The
second difficulty is that the statistics of the CHORUS and NOMAD experiments
will not be high enough to distinguish between models with close $\kappa$
values (for example $3 \over 8$ and $1 \over 2$). Hence, if these experiments
do discover $\numt$ oscillations, it will be very important to do further
experiments with higher statistics, such as the Fermilab proposal P803[16]. For
example, if  $\kappa = {1 \over 2}$, P803 will be able to determine the value
of $\sin^2 2\thetut$ to an accuracy of about 10\%, assuming that $\Delta m^2$
is large enough.

Finally, even if the assumptions made in this paper are incorrect, grand
unified theories are still expected to yield relations of the form of eq.
(1.1): $\thetut = \kappa |\vcb| \eta$. The difference is that, in this more
general case, $\kappa$ need not take a special value corresponding to a group
theory Clebsch. For example it might be a linear combination of several
Clebschs. Nevertheless, the expectation is that $\kappa \approx 1$. In
principle $\eta$ could differ from the values shown in figure 1. In practice,
theories with perturbative couplings do not give results very different from
those of the MSSM shown in figure 1. Hence we conclude that CHORUS and NOMAD
will probe precisely the range of $\thetut$ of interest to grand unified
theories. A null result by P803 would imply that, within the context of grand
unified theories, $m_{\nu_\tau} \leq 3$ eV, too small to be of much interest
for dark matter.

\newpage
\hspace{2in} Figure Captions
\vskip 0.5in
 Figure 1: A plot of the renormalization correction factor $\eta$
 versus $\tan \beta$ for $\alphas(\mz) = .110,\, .118,\, \mbox{and} \, .126$.
 On the solid (dashed) [dotted] curve the $\overline{\mbox{MS}}$ values of
 the running b quark mass is $m_b(m_b)=4.25\,(4.35)\,[4.15]$ GeV.
\vskip 0.5in
 Figure 2: The $\Delta m^2$ vs. $\sin^2 2\thetut$ plot including the current
(FNAL 531) and future (NOMAD and CHORUS)  expected limits, and the primary
predictions of this paper (for $\kappa = {1 \over 2}, 1$), with error bars.
\end{document}